\documentclass[11pt,a4paper]{article}
\pdfoutput=1

\usepackage{graphicx}
\usepackage{caption}
\usepackage{subcaption}
\usepackage{bm}
\usepackage{amsmath}
\usepackage{amssymb,xfrac}
\usepackage[bbgreekl]{mathbbol}
\usepackage{cite,float}
\usepackage{enumitem}
\usepackage{xcolor}

\newcounter{prop}
\newcommand{\pn}[1]{\vspace{.2cm}
\noindent\textbf{Pn}
\stepcounter{prop}$\bm{\arabic{prop}}$: #1}

\newcounter{prof}
\newcommand{\pf}[1]{\vspace{.1cm}
\noindent\textbf{Pf}
\stepcounter{prof}$\bm{\arabic{prof}}$: #1
\vspace{.2cm}}

\newcommand\shrink{\hspace{.4em}}

\newcommand\undermat[2]{%
  \makebox[0pt][l]{$\smash{\underbrace{\phantom{%
    \begin{matrix}#2\end{matrix}}}_{\text{$#1$}}}$}#2}

\def\be{\begin{equation}}
\def\ee{\end{equation}}
\def\bea{\begin{eqnarray}}
\def\eea{\end{eqnarray}}

\def\rd{\mathrm{d}}
\def\ri{\mathrm{i}}
\def\I{\mathrm{I}}

\begin{document}

\begin{center}
\noindent {\large\textbf{On U(1) Gauge Theory Transfer-Matrix in Fourier Basis
}}
\\[1\baselineskip]
Narges Vadood~\footnote{n.vadood@alzahra.ac.ir} ~~~~~and~~~~~Amir H. Fatollahi~\footnote{Corresponding Author: fath@alzahra.ac.ir}
\\[1\baselineskip]
\textit{Department of Physics, Alzahra University, Tehran 1993891167, Iran}
\\[1\baselineskip]
\begin{abstract}
\noindent
The properties of the transfer-matrix of U(1) lattice gauge theory 
in the Fourier basis are explored. 
Among other statements it is shown:
1) the transfer-matrix is block-diagonal,
2) all consisting vectors of a block are known based on 
an arbitrary block vector, 3) the ground-state belongs to 
the zero-mode's block. The emergence of maximum-points in 
matrix-elements as functions of the gauge coupling is clarified. 
Based on explicit expressions
for the matrix-elements we present numerical results as tests of our statements.
\end{abstract}
\end{center}



\section{Introduction}

Currently the numerical studies of gauge theories
in the non-perturabative regime are mainly based on the
lattice formulation of these theories \cite{wilson,kogut,rothe,detar}.
The theoretical \cite{banks,savit,guth,spencer,jaffe} as well as 
the numerical \cite{arnold,langfeld,creutz,nauen,bha,moria,degran}
studies suggest that the compact 4D U(1) gauge theory 
possesses two different phases, the so-called Coulomb and confined ones.
Different studies suggest that the phase transition 
occurs at a critical coupling of order 
unity \cite{banks,savit,guth,spencer,jaffe,arnold,langfeld,creutz,
nauen,bha,moria,degran}. 

The advantages of Fourier transform of lattice gauge variables have 
already been shown in the so-called dual formulation of 
the theory \cite{banks, savit}, by which an insightful 
picture for the phases of U(1) theory is provided. 
Accordingly, in a certain small coupling 
limit known as Villain approximation \cite{villain} and
via the Fourier basis, the partition function by the U(1) model looks like 
the one of monopoles or circulating-monopoles in 3D and 4D cases, 
respectively \cite{banks}. 
As a consequence, depending on temperature the system may
exhibit different phases based on the spatial extent of the electric field out of 
an electric charge \cite{banks}. 
As other studies based on the dual formulation see \cite{d1,d2,d3}.

In the present work the main concern is the transfer-matrix 
on its own as the basic tool to define the quantum theory on a Euclidean lattice 
\cite{wilson, kogut}. In particular, regarding the transfer-matrix 
in the Fourier basis, some mathematical statements are presented. 
The ultimate goal of studies in this direction is 
to provide more detailed information about the transfer-matrix 
based on the first principles of lattice gauge theories, leading to the better understanding 
of the energy spectrum of these models. Except the asymptotic behaviors of the 
matrix-elements, the statements are obtained with no use of approximation based on the 
value of gauge coupling, and are valid in any lattice size and dimension.
The issues to be addressed include: the block-diagonal nature of matrix in the Fourier basis, 
the consisting vectors of each block, the small and large coupling limits of the matrix elements,
and the block to which the ground-state belongs. Also the emergence of maximum-points in the matrix-elements as functions of gauge coupling is discussed and clarified. 
Based on the explicit form of matrix-elements for the 3D case 
as unconstrained summations, some pieces of numerical results are presented as 
the prompt test of the statements, all confirming the announced results. 

The organization of the rest of the paper is as follows. In Sec.~2 the 
elements of the transfer-matrix are derived in the Fourier basis. In Sec.~3 the properties of the 
transfer-matrix and its elements in the Fourier basis are explored and formulated
in six propositions. Also based on the properties a simplified expression for the matrix-elements are
obtained which is more convenient for numerical purposes. 
In Sec.~4 the numerical results based on the obtained expression is presented as 
a demonstration for the practical use of the obtained expression as well as 
the tests of the statements. Sec.~5 is devoted to the conclusions.

\section{Transfer-Matrix in Fourier Basis}

The matrix element of the transfer-matrix $\widehat{V}$
between two adjacent times $n_0$ and $n_0+1$ is given by \cite{wipf}
\begin{align}\label{1}
\langle n_0+1 | \widehat{V} | n_0 \rangle = \mathcal{A}~
e^{ S_E(n_0,n_0+1)}
\end{align}
in which $\mathcal{A}$ is inserted to fix the normalization, and 
$S_E(n_0,n_0+1)$ is the Euclidean action symmetrized 
in variables of two adjacent times. 
Following \cite{crue,luscher,seiler} here we work in the 
temporal gauge $A^{0}\equiv 0$, in which the transfer-matrix gets a particularly simple form. 
It is convenient to replace the gauge variables at adjacent times 
$A_{n_0}^{(\bm{r},i)}$ and $A_{n_0+1}^{(\bm{r},i)}$ on spatial link $(\bm{r},i)$ 
by the new angle variables \cite{wilson}
\begin{align}\label{2}
\begin{aligned}\theta^{(\bm{r},i)}&=a\,g\,A_{n_0}^{(\bm{r},i)}\cr
\theta'^{(\bm{r},i)}&=a\,g\,A_{n_0+1}^{(\bm{r},i)}
\end{aligned}
\end{align}
both taking values in $[-\pi,\pi]$ \cite{wilson}. In Eq.~2 
$a$ and $g$ are the lattice spacing parameter and the gauge coupling, respectively.
The symmetrized Euclidean action in Eq.~1 for pure U(1) theory 
in temporal gauge on a lattice with $d$ spatial dimensions 
is explicitly given by \cite{luscher,seiler}
\begin{align}\label{3}
S_E(n_0,n_0+1)=- \frac{1}{2\, g^2} \sum_{\bm{r}}\sum_{i\neq j =1}^d 
& \left[2-\cos\big(\theta^{(\bm{r},i)}+\theta^{(\bm{r}+\widehat{i},j)} 
-\theta^{(\bm{r}+\widehat{j},i)}-\theta^{(\bm{r},j)}\big) \right. 
\cr
& \left.
-\cos\big(\theta'^{(\bm{r}, i)}+\theta'^{(\bm{r}+\widehat{i},j)} 
-\theta'^{(\bm{r}+\widehat{j},i)}-\theta'^{(\bm{r},j)}\big)\right]\cr
-\frac{1}{g^2} &\sum_{\bm{r}}\sum_{i=1}^d 
\left[1-\cos\big(\theta^{(\bm{r},i)}-\theta'^{(\bm{r},i)}\big)\right]
\end{align}
in which $\widehat{i}$ is the unit-vector along the spatial direction $i$.
For a spatial lattice with $N_\mathrm{P}$ plaquettes and
$N_\mathrm{L}$ links, it is convenient to define the 
plaquette-link matrix $\bm{M}$ of dimension $N_\mathrm{P}\times N_\mathrm{L}$
given by
\begin{align}\label{4}
M^p_{~l}=\begin{cases}
                     +1,& \mbox{link $l=(\bm{r},i)$ belongs to oriented plaquette $p$ } \\
                    -1,& \mbox{link $l=(\bm{r},-i)$ belongs to oriented plaquette $p$} \\
                     ~~0,& \mbox{otherwise.}
        \end{cases}
\end{align}
\begin{figure}[t]
		\includegraphics[scale=2.2]{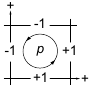}
\vskip -2cm  \caption{\small Graphical representation of definition (\ref{4}). }
\vskip 1.4cm
\end{figure}
In Fig.~1 the above definition is presented graphically. 
An explicit example for definition (\ref{4}) in $d=2$ case will be given later.
In terms of this matrix, 
labeling links as $l=(\bm{r},i)$ and plaquettes as $p$,
the action (\ref{3}) comes to the following form
\begin{align}\label{5}
S_E(n_0,n_0+1)=-\frac{1}{2g^2} \sum_p &
\left[2-\cos\big(M^p_{~l}\,\theta^l\big)
-\cos\big(M^p_{~l}\,\theta'^l\big)\right]
\cr &
-\frac{1}{g^2} \sum_l
\left[1-\cos\big(\theta^l-\theta'^l \big)\right]
\end{align}
in which summation over repeated indices is understood. Using 
\begin{align}\label{6}
\gamma=\frac{1}{g^2}
\end{align}
and by Eq.~(\ref{5}) the matrix-element (\ref{1}) may be written as
\begin{align}\label{7}
\langle \bm{\theta'} |\widehat{V} | \bm{\theta}\rangle =~ \mathcal{A}~ &
\prod_p \exp\!\left\{-\frac{\gamma}{2}\left[2-\cos\big(M^p_{~l}\,\theta^l\big)
-\cos\big(M^p_{~l}\,\theta'^l\big)\right]\right\}\cr
&\times\prod_{l}\exp\!\left\{-\gamma\Big[1-\cos\big(\theta^l-\theta'^l\big)\Big]\right\}
\end{align}
According to $\widehat{V}=\exp(-a\,\widehat{H})$, the eigenvalues 
$E_i$ of the Hamiltonian and $v_i$ of the transfer-matrix are related by
\begin{align}\label{8}
E_i=-\frac{1}{a}\ln v_i
\end{align}
As the main tool of this work, we formulate the theory in the 
Fourier basis $| k_l \rangle$, which is related to the compact $\theta$-basis by
\begin{align}\label{9}
\langle\theta^{l'}|k_{l} \rangle&=\frac{\delta^{l'}_{~l}}{\sqrt{2\,\pi}}
\,\exp(\ri\,k_l\,\theta^l)
,~~~~~ 
k_l=0,\pm1,\pm2,\cdots
\end{align}
Using the expansion
\begin{align}\label{10}
\exp(x\,\cos\phi)=\sum_n\I_n(x)\,\exp(\ri\,n\,\phi)
\end{align}
for the modified Bessel functions $\I_n(x)$ of the first kind
and the relation 
\begin{align}\label{11}
\int_{-\pi}^\pi\rd\theta\,\exp(\ri\,n\,\theta)=2\pi\,\delta_{n,0}
\end{align}
one directly finds the matrix elements of $\widehat{V}$ in the field Fourier basis 
\begin{align}\label{12}
\langle\bm{k}'|\widehat{V}|\bm{k}\rangle=\mathcal{A}
\, e^{-\gamma(N_\mathrm{P}+N_\mathrm{L})}
\,(2\pi)^{N_\mathrm{L}}
\sum_{\{n_p\}}\sum_{\{n'_p\}}\prod_{p}
\I_{n_p}\!\left(\frac{\gamma}{2}\right)\,\I_{n'_p}\!\left(\frac{\gamma}{2}\right)
\;\prod_{l} \I_{m_l}\!(\gamma)\,
\delta_{m_l,m'_l},
\end{align}
with 
\begin{align}\label{12.5}
m_l=k_l+\sum_p n_p\,M^p_{~l},~~~~~m'_l=\,k'_l-\sum_p n'_p\,M^p_{~l}.
\end{align}
\noindent where $n_p$, $n'_p$, $m_l$ and $m'_l$ are all integer-valued.

\section{Properties of Fourier Matrix-Element}

The matrix-element (\ref{12}) in the Fourier basis is the basis expression based on 
it in the following some propositions (by \textbf{Pn}'s) and their
proofs (by \textbf{Pf}'s) are presented:

\pn{For every matrix element we have the following properties:\\
1) non-negativity: $\langle\bm{k}'|\widehat{V}|\bm{k}\rangle \geq 0$,
2) symmetry: 
$\langle\bm{k}'|\widehat{V}|\bm{k}\rangle=\langle\bm{k}|\widehat{V}|\bm{k}'\rangle$,
3) reflectivity: 
$\langle\bm{k}'|\widehat{V}|\bm{k}\rangle=\langle-\bm{k}'|\widehat{V}|-\bm{k}\rangle$.
}

\pf{All of the above properties are evident using the properties
$\I_n(x)\geq 0$ and $\I_n(x)=\I_{-n}(x)$,
and appropriate sign-changes of the indices $n_p$, $n'_p$ and $m_l$.}

It is obvious that not only all matrix elements are non-negative, but also 
each term is so in the sum (\ref{12}). The vanishing of a matrix element means that 
the difference $\bm{k}'-\bm{k}$ can not
satisfy all the Kronecker $\delta$'s in Eq.~(\ref{12}) for any set of 
integers $\{n_p, n'_p\}$. 

\pn{All diagonal elements are non-zero: 
$\langle \bm{k}|\widehat{V}|\bm{k}\rangle\neq 0$.}

\pf{It is easy to see that there are always surviving terms for $\bm{k}'=\bm{k}$ 
in Eq.~(\ref{12}). On the diagonal $\bm{k}=\bm{k}'$, 
setting all $n_p+n'_p=0$ is enough to satisfy
all $\delta$'s in Eq.~(\ref{12}), leading to non-vanishing positive terms.}

\noindent In fact, for satisfying $\delta$'s in Eq.~(\ref{12}) with $\bm{k}=\bm{k}'$,
it is sufficient to set $n_p+n'_p={n}^0_p$ with the condition 
$\sum_p {n}^0_p M^p_{~l}=0$, presented in the vector notation as
\begin{align}\label{13}
\bm{n}^{\bm{0}}\cdot \bm{M}=\bm{0}
\end{align}
Later we will give the general form of the non-zero elements based on the 
$\bm{n^0}$ vectors.

\pn{Transitivity: If $\langle \bm{k}|\widehat{V}|\bm{k}'\rangle \neq 0$ and 
$\langle \bm{k}'|\widehat{V}|\bm{k}''\rangle \neq 0$, then
$\langle \bm{k}|\widehat{V}|\bm{k}''\rangle \neq 0$.}

\pf{This simply follows by two successive uses of the $\delta$'s 
in Eq.~(\ref{12}). }

By \textbf{Pn~2~\&~3}, having a non-zero matrix element is an 
\textit{equivalence relation}, by which the set of all 
$\bm{k}$'s is partitioned into 
\textit{equivalence classes}. Later by explicit examples we will see
that there is more than one class (in fact, an infinite number of classes) 
even for a finite size lattice. As a consequence, 
the transfer-matrix $\widehat{V}$ appears in the
block-diagonal form based on the classes, with all elements of each block 
being non-zero. The remarkable fact is that, 
given by a Fourier mode $\bm{k}$ one can simply construct
all of its co-blocks. 
This is simply done by setting $n_p+n'_p=q_p+{n}^0_p$,
in which $q_p$'s are arbitrary. Then by using $\sum_p {n}^0_p M^p_{~l}=0$ 
another mode in the class is obtained as $k^{\{q\}}_l = k_l‎ + ‎\sum_p q_p \, ‎M^p_{~l}$,‎
presented in the vector notation by 
\begin{align}\label{14}
\bm{k}^{\bm{q}}=\bm{k} + \bm{q}\cdot \bm{M}
\end{align}
It is obvious by definition (\ref{12.5}) that the two modes $\bm{k}$ and $\bm{k}^{\bm{q}}$ satisfy all $\delta$'s in Eq.~(\ref{12}). 
Also if $\bm{q}$ satisfy the condition (\ref{13}), 
the two modes are the same ($\bm{k}^{\bm{q}}=\bm{k}$). 
For two modes related by Eq.~(\ref{14}) the non-zero matrix-element simply gets the form
\begin{align}\label{15}
\langle\bm{k}^{\bm{q}}|\widehat{V}|\bm{k}\rangle=\mathcal{A}
\, e^{-\gamma(N_\mathrm{P}+N_\mathrm{L})}
\, (2\pi)^{N_\mathrm{L}}   &
\sum_{\{{n}^0_p\}}  
\delta\Big(\sum_p {n}^0_p M^p_{~l}\Big) \sum_{\{n_p\}}
\cr
&\prod_{p}
\I_{n_p}\!\left(\frac{\gamma}{2}\right)\I_{q_p+{n}^0_p-n_p}
\!\left(\frac{\gamma}{2}\right)
\prod_{l} \I_{k_l+\sum_p \!\! n_p M^p_{~l}}\!(\gamma)
\end{align}
The important fact is that the allowed ${n}^0_p$'s are not depending 
on $\bm{k}$, 
but only on the matrix $\bm{M}$. As an instructive example, let us consider 
the case of a 2d periodic spatial lattice, for which we later explicitly find that 
the sub-space of the vectors $\bm{n^0}$ satisfying condition (\ref{13}) 
is one-dimensional with the general form
\begin{align}\label{16}
\bm{n^0}=n^0 \underbrace{(1,1,\cdots,1)}_{N_\mathrm{P}} = n^0\,\bm{s}
\end{align}
For periodic lattices with $d=2$ the matrix-element (\ref{15}) gets the form
\begin{align}\label{17}
\langle\bm{k}^{\bm{q}}|\widehat{V}|\bm{k}\rangle=\mathcal{A}\,
 e^{-\gamma(N_\mathrm{P}+N_\mathrm{L}) }  (2\pi)^{N_\mathrm{L}}
\sum_{{n}^0}   \sum_{\{n_p\}}
& \prod_{p}\I_{n_p}\!\left(\frac{\gamma}{2}\right)\I_{q_p+{n}^0-n_p}
\!\left(\frac{\gamma}{2}\right) 
\cr &
\prod_{l} \I_{k_l+\sum_p \!\! n_p M^p_{~l}}\!(\gamma)
\end{align}
This expression, with no restriction on summations, 
is quite adequate for numerical purposes and will be used later. 

\pn{Each block of $\widehat{V}$ is infinite dimensional}.

\pf{This simply follows  by the infinite possible choices for the integer
sets $\{q_p\}$'s.}

\noindent For definiteness, throughout this work we consider 
the normalization $\mathcal{A}$
to be constant (\textit{i.e.} independent of $g$);
for other choices and their consequences see \cite{spchfath}. 
The limit $\gamma\ll 1$ (large coupling limit $g\gg 1$) of the 
matrix elements is obtained easily by the expansion of exponentials
in Eq.~(\ref{7}), by which in the lowest orders one finds
\begin{align}\label{18}
\langle\bm{k}'|\widehat{V}|\bm{k}\rangle&=\mathcal{A}
\, e^{-\gamma(N_\mathrm{P}+N_\mathrm{L})}
\,(2\pi)^{N_\mathrm{L}}\,\bigg\{
\prod_{l}\delta(k_l)\,\delta(k'_l)\cr
&+\frac{\gamma}{4}\sum_{p}
\bigg[\prod_{l}\delta\left(k_l+M^p_{~l}\right)\delta(k'_l)
+\prod_{l}\delta\left(k_l-M^p_{~l}\right)\delta(k'_l)\cr
&~~~~~~+\prod_{l}\delta(k_l)\delta\left(k'_l-M^p_{~l}\right)
+\prod_{l}\delta(k_l)\delta\left(k'_l+M^p_{~l}\right)\bigg]\cr
& + \frac{\gamma}{2}\bigg[\prod_{l}\delta(k_l+1)\,\delta(k'_l-1)
+\prod_{l}\delta(k_l-1)\,\delta(k'_l+1)\bigg]+O(\gamma^2)\bigg\}
\end{align}
This leads to the next important proposition:

\pn{Provided that the ground-state is unique, it belongs to the
$\bm{k}=\mathbf{0}$ block.}

\pf{According to expansion (\ref{18}), in the $\gamma\to 0$ limit
all the elements of $\widehat{V}$ are approaching zero, except the 
diagonal element $V_{\mathbf{00}}=\langle \mathbf{0} |\widehat{V}|\mathbf{0}\rangle$.
By the relation (\ref{8}) between energy eigenvalues and $\widehat{V}$-eigenvalues, all energies are going to infinity in limit $\gamma\to 0$ except 
the one in $V_{\mathbf{00}}$'s block, appearing as the lowest energy. 
Since lowering the coupling (increasing $\gamma$) does not cause a mixing among the blocks, by the uniqueness assumption, 
the ground-state belongs to the $\bm{k}=\mathbf{0}$'s block at any coupling.}

The other interesting limit is at $\gamma\to \infty$ ($g\to 0$), which is expected to recover the ordinary formulation of the gauge theory in the continuum. This limit can be reached by using 
the asymptotic behavior of Bessel functions for large arguments. They
read in the saddle-point approximation
\begin{align}\label{19}
\I_n(x)\simeq \frac{e^x}{\sqrt{2\pi x}}\,
e^{-n^2/2x+1/8x}\,\big(1+\mathrm{O}(1/x^2)\big)
,~~~~ x\to \infty 
\end{align}
by which in the $\gamma\to\infty$ limit the terms in the 
matrix-element (\ref{17}) can be treated as Gaussian integrals, leading to the asymptotic behavior
\begin{align}\label{20}
\langle\bm{k}^{\bm{q}}|\widehat{V}|\bm{k}\rangle\simeq
\mathcal{A}~
\frac{(2\pi)^{\frac{1}{2}(N_\mathrm{L}+N_\mathrm{P})}}
{\pi^{N_\mathrm{P}-\frac{1}{2}}\sqrt{\det \bm{C}}}
\frac{1}{\sqrt{\bm{s}^T\!\bm{D}\bm{s}}}~
\frac{e^{-B(\bm{k},\bm{q})/\gamma}}{\gamma^{\frac{1}{2}(N_\mathrm{P}+N_{\mathrm{L}}-1)}}\,
e^{b/\gamma}
\end{align}
in which $b=(4N_\mathrm{P}+N_\mathrm{L})/8$, and  
$B(\bm{k},\bm{q})$, in terms of the symmetric matrices 
$\bm{C}$, $\bm{D}$ and $\bm{F}$, is
\begin{align}\label{21}
B(\bm{k},\bm{q})=
\bm{q}^T\! \bm{D}\bm{q}+\frac{1}{2}\bm{k}^T\!\bm{F}\bm{k}
+2\bm{q}^T\!\bm{C}^{-1}\!\bm{M}\bm{k}
-\frac{\big[\bm{s}^T\!(\bm{D}\bm{q}+\bm{C}^{-1}\!\bm{M}\bm{k})\big] ^2}
{\bm{s}^T\!\bm{D}\bm{s}}
\end{align}
with $\bm{s}$ given in Eq.~(\ref{16}), and 
\begin{align}\label{22}
\bm{C}&=4\,\mathbb{1}_{N_\mathrm{P}}+\bm{M}\bm{M}^T\!,~~
\bm{D}=\mathbb{1}_{N_\mathrm{P}}- 2\,\bm{C}^{-1}\!,~
~\bm{F}=\mathbb{1}_{N_\mathrm{L}}-\bm{M}^T\!\bm{C}^{-1}\!\bm{M}
\end{align}
in which $\mathbb{1}_{N}$ is the identity matrix of dimension $N$.
For spatial lattices with dimensions larger than two, taking the dimension
of sub-space of vectors $\bm{n^0}$ as $N_0$, the asymptotic 
behavior again can be obtained as 
$\langle\bm{k}^{\bm{q}}|\widehat{V}|\bm{k}\rangle\propto 
e^{\alpha/\gamma}/\gamma^{\frac{1}{2}(N_\mathrm{P}+N_{\mathrm{L}}-N_0)}$, by which or by (\ref{20}) 
the matrix-elements tend to zero by $\gamma\to \infty$ in any dimension. 
On the other hand by expansion (\ref{18}),
we already know that only $V_{\mathbf{00}}$ may survive in the limit
$\gamma\to 0$. An immediate conclusion is:

\pn{
Except perhaps $V_{\mathbf{00}}$, all non-zero matrix-elements are to develop maximum.} 

\pf{As by \textbf{Pn~1} all non-zero matrix elements are positive, for the mentioned elements 
the increasing behavior at small $\gamma$ and the decreasing one at large $\gamma$ 
are to be connected through at least one maximum.}

\noindent Our numerical results demonstrate clearly 
the appearance of precisely one maximum. 
The existence of the maximum in matrix elements of $\widehat{V}$ 
has particularly important consequences on the phases of the model; an
issue that we do not discuss further and leave for later works.

\section{Numerical Results}
In this section examples of numerical results are presented based on the expressions obtained in the Fourier basis. The aim for presenting the numerical results 
is twofold. First to show how the final expressions in the Fourier basis, such as Eq.~(\ref{17})
can be used practically for generating numerical results. Second is to provide the tests
for the statements presented in previous section, including the vectors 
belonging to a common block, and 
the appearances of maximum-points in the matrix elements.

\begin{figure}[t]
	\begin{center}
		\includegraphics[scale=.26]{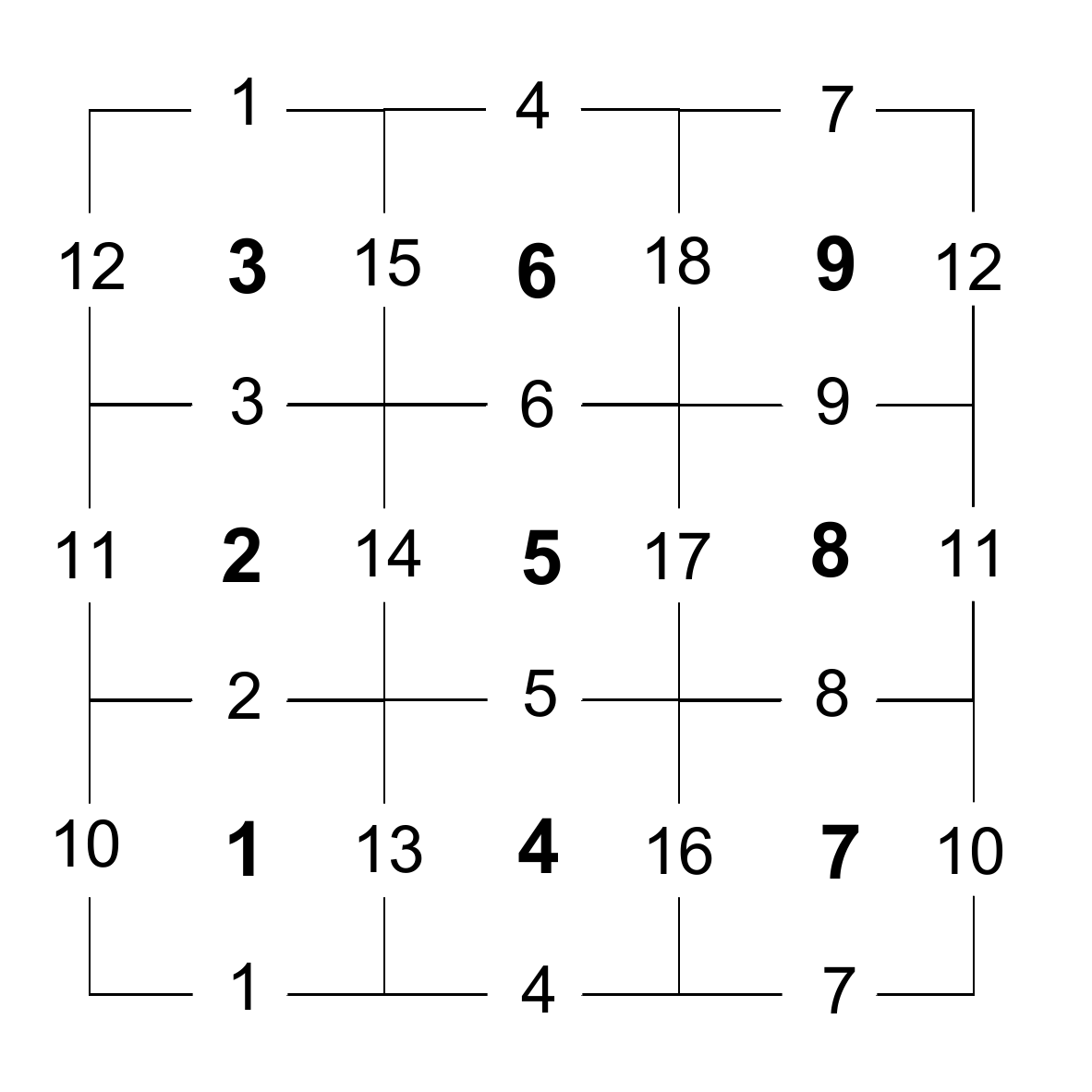}
	\end{center}
\vskip -.6cm   \caption{\small The numbering of links and plaquettes for the $3\times 3$
\textit{periodic} lattice used in representation of matrix $\bm{M}$ (\ref{26}). }
\end{figure} 
To proceed let us have an explicit representation of 
the plaquette-link matrix $\bm{M}$. In the following we 
consider a lattice with two spatial dimensions $d=2$. 
For a 2-dim cubic periodic lattice with $N_s$ sites 
in each direction, it is convenient to define the $N_s\times N_s$  translation-matrix $\bm{T}$ by
\begin{align}\label{23}
T_{ab}= \delta_{ab}-\delta_{a+1,b}-\delta_{a,N_s}\,\delta_{b1},~~~~~
a,b=1,\cdots,N_s
\end{align}
For $N_s=3$ the explicit form of $\bm{T}$ is
\begin{align}\label{24}
\bm{T}=\begin{pmatrix}
+ & - & 0 \cr
0 & + & -  \cr
- & 0 & + 
\end{pmatrix}
\end{align}
For $N_s$ sites in each direction of a 2-dim periodic cubic lattice 
there are $N_\mathrm{P}=N_s^2$ plaquettes and 
$N_\mathrm{L}=2N_s^2$ links.
Then, by the numbering of plaquettes and links as shown in Fig.~2, it is easy to check
that the matrix $\bm{M}$ can be constructed by gluing the two $N_s^2\times N_s^2$ matrices next to each other, as follows
\begin{align}\label{25}
\bm{M}=\left(\begin{array}{c|c}
  & \cr
 \mathbb{1}_{N_s}\otimes \bm{T} ~ &   - \bm{T}\otimes \mathbb{1}_{N_s} \cr 
  & 
\end{array}\right)
\end{align}
\noindent By construction, the matrix $\bm{M}$ is the $N_s^2\times 2N_s^2$ dimensional, as it should. For $N_s=3$ the matrix gets the form
\begin{align}\label{26}
\bm{M}=\left(\begin{array}{c@{\shrink}c@{\shrink}c@{\shrink}c@{\shrink}
c@{\shrink}c@{\shrink}c@{\shrink}c@{\shrink}c@{\shrink\shrink}
c@{\shrink}c@{\shrink}c@{\shrink}c@{\shrink}c@{\shrink}c@{\shrink}
c@{\shrink}c@{\shrink}c}
 + & - & 0 & 0 & 0 & 0 & 0 & 0 & 0 & - & 0 & 0 & + & 0 & 0 & 0 & 0 & 0 \\
 0 & + & - & 0 & 0 & 0 & 0 & 0 & 0 & 0 & - & 0 & 0 & + & 0 & 0 & 0 & 0 \\
 - & 0 & + & 0 & 0 & 0 & 0 & 0 & 0 & 0 & 0 & - & 0 & 0 & + & 0 & 0 & 0 \\
 0 & 0 & 0 & + & - & 0 & 0 & 0 & 0 & 0 & 0 & 0 & - & 0 & 0 & + & 0 & 0 \\
 0 & 0 & 0 & 0 & + & - & 0 & 0 & 0 & 0 & 0 & 0 & 0 & - & 0 & 0 & + & 0 \\
 0 & 0 & 0 & - & 0 & + & 0 & 0 & 0 & 0 & 0 & 0 & 0 & 0 & - & 0 & 0 & + \\
 0 & 0 & 0 & 0 & 0 & 0 & + & - & 0 & + & 0 & 0 & 0 & 0 & 0 & - & 0 & 0 \\
 0 & 0 & 0 & 0 & 0 & 0 & 0 & + & - & 0 & + & 0 & 0 & 0 & 0 & 0 & - & 0 \\
\undermat{\mathbb{1}_{3}\otimes \bm{T}}{ 0 & 0 & 0 & 0 & 0 & 0 & -
&} 0 &  + & 
\undermat{-\bm{T}\otimes \mathbb{1}_{3}}{ 0 & 0 & + & 0 & 0 & 0 & 0  & } 0 & -  \\
\end{array}
\right)
\end{align}
\vskip .5cm

\noindent As announced before Eq.~(\ref{16}), it is an obvious consequence
of this explicit form of $\bm{M}$ that the sub-space of $\bm{n^0}$ of Eq.~(\ref{16}) is one-dimensional. As two vectors in an equivalence class consider 
the followings
\begin{align}\label{27}
|\bm{0}\rangle &\to \bm{k}=(0,0,\cdots,0)
\\  \label{28}
|\bm{1}\rangle &\to \bm{k^{q_1}}=\bm{q_1}\cdot\bm{M}
\end{align} 
in which $\bm{q_1}=(1,0,\cdots,0)$ with $N_\mathrm{P}=N_s^2$ elements. 
By expansion (\ref{18}) and \textbf{Pn~5}, the non-vanishing elements $V_{\bm{00}}$, $V_{\bm{01}}=V_{\bm{10}}$, and $V_{\bm{11}}$
belong to the ground-state's block.
To see how a vanishing element may occur, consider
\begin{align}\label{29}
|\bm{1'}\rangle &\to \bm{k_{1'}}=(1,0,0,\cdots,0)\\
\label{30}
|\bm{1''}\rangle &\to \bm{k_{1''}}=(0,1,0,\cdots,0)\\
\label{31}
|\bm{2}\rangle &\to \bm{k_2}=(2,0,0,\cdots,0)
\end{align}
By an explicit representation like matrix (\ref{26}), it is seen that any pair of 
the above vectors can not satisfy Eq.~(\ref{14}), leading to vanishing elements $V_{\bm{1'1''}}=V_{\bm{1''2}}=V_{\bm{1'2}}=0$. In 
other words, by the given representation for $\bm{M}$ and by any pair of 
Eqs.~(\ref{29})-(\ref{31}) 
one can see there is no solution for the $\{n_p+n'_p\}$'s inside the $\delta$'s of 
Eq.~(\ref{12}). The same is true between each of Eqs.~(\ref{29})-(\ref{31})  and one of 
Eqs.~(\ref{27})-(\ref{28}). Hence, the five vectors (\ref{27})-(\ref{31}) belong to four different blocks. 

Using the explicit form of $\bm{M}$, the expression (\ref{17}) with
summations on $N_s^2+1$ integers in 2-dim case
is quite adequate for numerical purposes. In the following to provide 
a prompt test for the announced results some pieces of numerical results 
are presented. The first issue in doing the summations is 
about a suitable choice of cut-off (upper limit) for sums. 
For small $\gamma$ limit we have 
\begin{align}\label{32}
\I_s(\gamma)\simeq \frac{1}{s!}\left(\frac{\gamma}{2}\right)^{|s|},~~~~ \gamma\ll |s|
\end{align}
by which for small arguments the Bessel's of low degrees are quite dominant. 
The subtle point is about large arguments, 
for which an initial guess of cut-off  is $s_*\simeq \sqrt{2}\,\gamma$,
at which by behavior (\ref{19}) we have $\I_{s_*}(\gamma) \propto 1/\sqrt{\gamma} $. 
However, in practice a lower cut-off is sufficient, as in the 
summations there are multiple of Bessel functions rather than a single one,
making the convergence to the desired significant digits faster. 
As examples in the vacuum class the numerical evaluations of elements $V_{\bm{00}}$, $V_{\bm{01}}$, and 
$V_{\bm{11}}$ of modes (\ref{27}) and (\ref{28}) are presented in
Fig.~3, by the choice $\mathcal{A}=1$ and 
for $3\times 3$ and $10\times 10$ lattices. 
To see how the statements work in the non-vacuum classes the 
elements $V_{\bm{1'1'}}=V_{\bm{1''1''}}$ and $V_{\bm{22}}$ 
are presented in Fig.~4; as mentioned 
$V_{\bm{1'1''}}=V_{\bm{1''2}}=V_{\bm{1'2}}=0$ since the modes belong to
different classes. The results are generated on a desktop PC in reasonable time. 
Also the evaluated elements confirm numerically the announced asymptotic behavior (\ref{20}). 
As expected by \textbf{Pn~6}, except $V_{\bm{00}}$ all other elements develop maximum.

\begin{figure}[t]
\centering
\begin{subfigure}{.5\textwidth}
\centering
		\includegraphics[width=.95\linewidth]{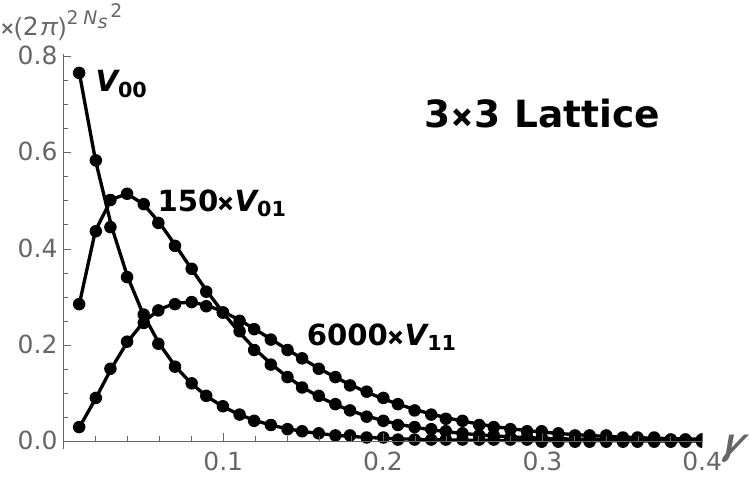}
\end{subfigure}%
\begin{subfigure}{.5\textwidth}
\centering
		\includegraphics[width=.95\linewidth]{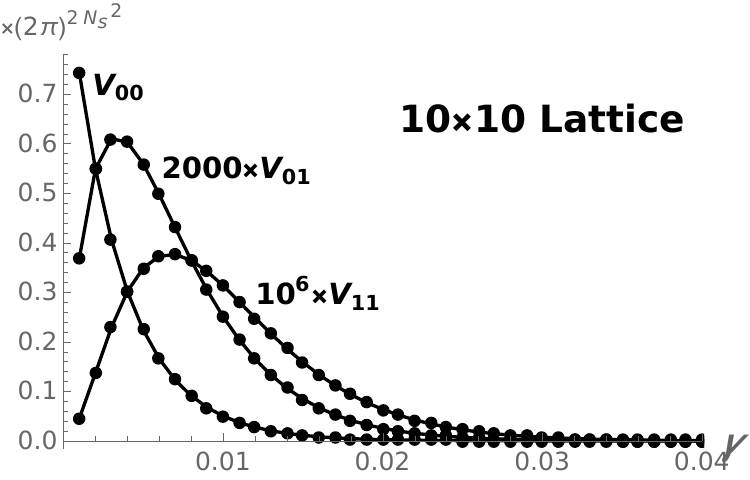}
\end{subfigure}
	\caption{The elements $V_{\bm{00}}$, $V_{\bm{01}}$, and $V_{\bm{11}}$
in vacuum class versus $\gamma$ by (\ref{17}) for normalization $\mathcal{A}=1$ for
2-dim lattices with $N_s=3$ and $N_s=10$. }
\end{figure}

\begin{figure}[t]
\centering
\begin{subfigure}{.5\textwidth}
\centering
		\includegraphics[width=.95\linewidth]{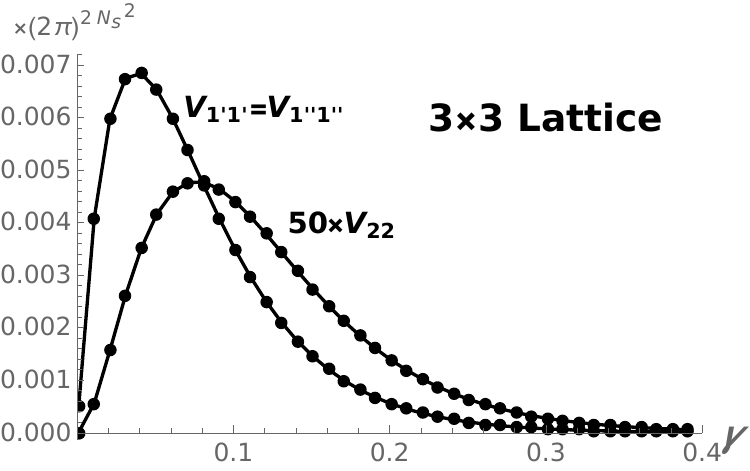}
\end{subfigure}%
\begin{subfigure}{.5\textwidth}
\centering
		\includegraphics[width=.95\linewidth]{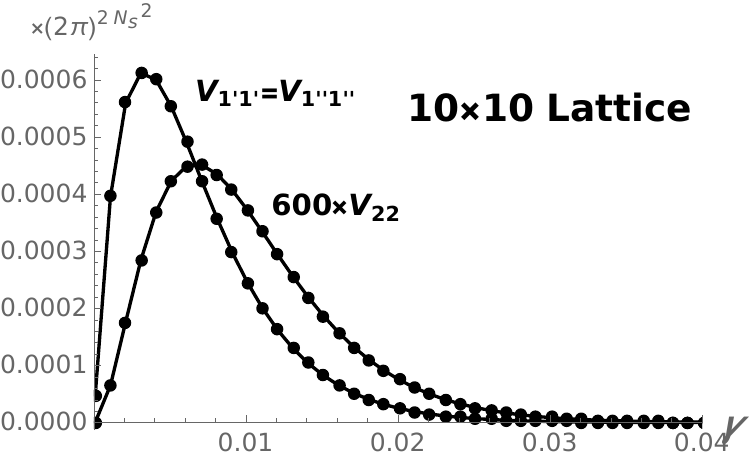}
\end{subfigure}
	\caption{The elements $V_{\bm{1'1'}}=V_{\bm{1''1''}}$ and 
$V_{\bm{22}}$ (all in different blocks) versus $\gamma$ by Eq.~(\ref{17}) for normalization $\mathcal{A}=1$ for
2-dim lattices with $N_s=3$ and $N_s=10$. }
\end{figure}

\section{Conclusion}

In summary, in the present work we explored the properties 
of the transfer matrix in the Fourier basis for the U(1) lattice gauge theory.
Regarding this matrix in the Fourier basis, some mathematical statements 
are presented, covering the issues: the block-diagonal nature of the matrix, 
the consisting vectors of each block, 
the small and large coupling limits of the matrix elements,
the block to which the ground-state belongs, and the 
appearance of maximum in the elements as functions of coupling. 
Based on the explicit form of matrix-elements Eq.~(\ref{17})
for the 3D case, samples of numerical results are presented all in agreement 
with the announced properties.
Apart from the asymptotic behaviors, the statements are obtained with no use 
of approximation based on the value of gauge coupling, and are valid in 
any lattice size and dimension.
It is a matter of importance to see how the formalism based on the
transfer-matrix in the Fourier basis regenerate the expected phase structures 
by the 3D and the 4D models \cite{banks}, specially in a quantitative way. 
In particular, 
one of the main questions in this direction is what features of the phase structure by the pure U(1) 
model on lattice would survive in the continuum limit.
This and further analytical and numerical results in this direction will be presented in future.


\vskip .6cm
\textbf{Acknowledgment:}
The authors are grateful to M. Khorrami for helpful discussions. 
This work is supported by the Research Council of the Alzahra University.

\end{document}